\documentclass[twocolumn]{aa}
\usepackage{graphicx}
\usepackage{txfonts}
\usepackage{xcolor}
\usepackage[normalem]{ulem}
\usepackage{mathtools}
\usepackage{adjustbox}

\newcommand{\Mm}{{\mathrm{\, Mm}}}
\newcommand{\kms}{{\mathrm{\, km \,s^{-1}}}}
\definecolor{rakesh}{rgb}{0., 0., 1.0}
\definecolor{vaibhav}{rgb}{0.55, 0.71, 0.0}
\definecolor{manuel}{rgb}{0.8, 0.0, 0.0}
\definecolor{dipankar}{rgb}{1, 0.49, 0.}

\authorrunning{Mazumder et al.}

\begin{document}

\title{Simultaneous longitudinal and transverse oscillations in filament threads after a failed eruption}

\author{Rakesh Mazumder \inst{1,2}
     \and
          Vaibhav Pant\inst{3}
          \and
          M. Luna\inst{4,5}
          \and
          Dipankar Banerjee\inst{1,2}
          }

\institute{Center of Excellence in Space Sciences India, Indian Institute of Science Education and Research Kolkata, Mohanpur 741246, West Bengal, India\\
              \email{rakesh@iiap.res.in}
         \and
             Indian Institute of Astrophysics, Bangalore-560 034, India
          \and
            Centre for mathematical Plasma Astrophysics, Department of Mathematics, KU Leuven, Celestijnenlaan 200B bus 2400, 3001 Leuven, Belgium\\
            \email{vaibhavpant55@gmail.com}
      \and
        Instituto de Astrof\'{\i}sica de Canarias, E-38205 La Laguna, Tenerife, Spain
         \and
             Departamento de Astrof\'{\i}sica, Universidad de La Laguna, E-38206 La Laguna, Tenerife, Spain
             }

   \date{Received --- ; accepted ---}




\abstract
{Longitudinal and transverse oscillations are frequently observed in the solar prominences and/or filaments. These oscillations are excited by a large scale shock wave, impulsive flares at one leg of the filament threads, or due to any low coronal eruptions. We report simultaneous longitudinal and transverse oscillations in the filament threads of a quiescent region filament. We observe a big filament in the north-west of the solar disk on $6^{th}$ July 2017. On $7^{th}$ July 2017 it starts rising around 13:00 UT. Then we observe a failed eruption and subsequently, the filament threads start to oscillate around 16 UT.}
{We observe both transverse and longitudinal oscillations in a filament thread. The oscillations damp down and filament threads almost disappear.}
{We place horizontal and vertical artificial slits on the filament threads to capture the longitudinal and transverse oscillations of the threads. Data from Atmospheric Imaging Assembly (AIA) onboard Solar Dynamics Observatory (SDO) is used to detect the oscillations. }
{We find the signatures of large amplitude longitudinal oscillations (LALOs). We also detect damping in LALOs. In one thread of the filament, we observe transverse oscillations (LATOs). Using the pendulum model we  estimate the lower limit of magnetic field strength and radius of curvature from the observed parameter of LALOs.}
{We show the co-existence of two different wave modes in the same filament threads. We estimate magnetic field from LALOs and hence suggest a possible range of the length of the filament threads using LATOs.}

\maketitle

\section{Introduction}
A filament is a cool structure in the solar corona appearing as a dark elongated feature in the solar disk. The same feature, when seen at the limb, appears bright and termed as prominence. The eruption of filament often produces Coronal Mass Ejection (CME) which has an adverse effect on space weather \citep{2000ApJ...537..503G,2003ApJ...586..562G,2004ApJ...614.1054J}. Thus, the study of the filament dynamics is quite relevant for the space weather objectives and forecasting.

The different modes of oscillation of the filaments provide valuable information on the prominence characteristics that are hard-to-measure by direct means. Prominence seismology combines theoretical modelling with observations to infer the local plasma conditions and magnetic properties \citep[see, e.g.,][]{2005LRSP....2....3N,2005ApJ...624L..57A,2011SSRv..158..169A,2012ASSP...33..159A}.
According to \citet{2002SoPh..206...45O} filament oscillations can be roughly divided into two categories: large-amplitude oscillations (LAOs) and small amplitude-oscillations. Small-amplitude oscillations in the filament have velocities less than 2-3
$\kms$ and localized in a small portion of the filament. Small amplitude oscillations reveal the local and small-scale properties of the plasma. In contrast, LAOs are oscillation with velocities larger than 10 $\kms$ and they disturb a large portion of the filament if not the whole filament. LAOs provide information on the global filament properties \citep[see,][for a discussion on the oscillation classification]{arregui2018,luna2018}. There are two types of LAOs, large-amplitude transverse oscillations (LATOs) and large-amplitude longitudinal oscillations (LALOs). The periods of LATOs are reported in the range of 6 minutes to 150 minutes \citep{2009SSRv..149..283T}, whereas the LALOs are within the range of 40 to 160 minutes. 

The observations reveal that LATOs and LALOs are excited from different sources. LATOs are generally triggered by distant flares, CMEs and Moreton waves \citep{1960PASP...72..357M,1966ZA.....63...78H,1966AJ.....71..197R,2002PASJ...54..481E,2004ApJ...608.1124O,2006A&A...449L..17I,2006ESASP.617E.141P,2007SoPh..246...89I,2008ApJ...685..629G,2011A&A...531A..53H,2013ApJ...773..166L,2018ApJ...860..113Z}. 
In some cases, LATOs are associated with filament eruptions  
 \citep{2006A&A...449L..17I,2007SoPh..246...89I,2008ApJ...680.1560P} .
In contrast, LALOs are generally excited by nearby microflares, small jets or partial eruption of the filament \citep{2003ApJ...584L.103J,2006SoPh..236...97J,2007A&A...471..295V,2011SoPh..269...83C,2012ApJ...760L..10L,2012A&A...542A..52Z,2014ApJ...785...79L,2017ApJ...842...27Z,luna2017}. However, \cite{2014ApJ...795..130S} have reported a LATO in a filament and LALO in another filament from the same shock wave. \cite{2015RAA....15.1713P}, \cite{2016SoPh..291.3303P} and \cite{2017ApJ...851...47Z} reported that both LALO and LATO  can be triggered simultaneously in the same filament by EUV wave from a nearby flare. Recently, \cite{2017Ap&SS.362..165C} reported that longitudinal oscillations along the filament threads can be sometimes (mis)interpreted as transverse oscillations. 


Explaining the nature of LALOs is challenging since strong restoring force is needed to create a huge acceleration. The energy associated is also enormous due to filament's huge mass and large velocity of the motions involved. \citet{2012ApJ...750L...1L} developed the so-called pendulum model based on the numerical simulations by \citet{2012ApJ...757...98L}. The authors found that gravity projected along the dipped magnetic field is the restoring force of the oscillations. \cite{2012A&A...542A..52Z} found a similar result by numerical means and compared the pendulum model with observations. The authors measured a LALOs of an active region prominence observed on 2007 February 6 with Hinode/SOT. Due to the high spatial resolution of the observation, the authors measured the curvature of the dipped field lines. They found a good agreement between the observations and the pendulum model. \citet{2013A&A...554A.124Z} performed a parametric theoretical study of LALOs. The simulations showed that, in LALOs, the gas-pressure gradients are small and do not influence the oscillation, indicating that the nature of LALOs is not magnetosonic. \citet{luna2016} found that the oscillations are dependent mainly on the radius of curvature of the magnetic field and no other details of the field line geometry. Additionally, \citet{luna2016a} found that there is no coupling between longitudinal motion and transverse motions in 2D numerical simulations. All these results indicate the robustness of the pendulum model for explanation and analysis of LALOs. More recently, \citet{luna2017} compared the magnetic field geometry inferred from LALO seismology with the field obtained from photospheric field extrapolation technique finding a very good agreement.
The pendulum model have been used to estimate the magnetic field and the radius of the curvature of the filament from LALOs \citep[i.e.,][]{2014ApJ...785...79L,2016SoPh..291.3303P}. 

Both LALOs and LATOs are damped oscillations \citep[see reviews by][]{2009SSRv..149..283T,arregui2018}.
 %
 %
 %
 Several mechanisms for strong damping in LALOs are 
 have been proposed (see review by 
 \cite{2009SSRv..149..283T}) but the matter is still under 
 debate. 
 Mass accretion by the cool prominence has been suggested for the strong damping of LALOs by
 \citep{2012ApJ...750L...1L,2014ApJ...785...79L,2016SoPh..291.3303P,2016ApJ...817..157L,2016A&A...591A.131R}. Similarly, \cite{2013A&A...554A.124Z} suggested mass drainage as another mechanism for the damping of LALOs.
 Additionally, \citet{2013A&A...554A.124Z} performed a parametric numerical study and found that the radiative cooling could explain the LALOs' damping.
 In contrast, LATOs have a magnetic origin and it is accepted that resonant absorption is the cause of their damping.
 %

There is an increasing number of publications on LAOs since the early observations at the beginning of the 20th century \citep[see review by][]{2009SSRv..149..283T}. \citet{luna2018} catalogued almost 200 oscillations in prominences with nearly half of the events being LAOs. The authors found that LAO events are very common with one LAO every two days on average on the visible side of the Sun. The large number of LAO events makes the prominence seismology a very powerful tool to infer the global characteristics of the filaments.

In this work, we report an interesting event in which we observed both LALOs and LATOs in filament threads followed by a failed filament eruption. Partial filament eruption leading to the transverse oscillations in the nearby coronal loops is reported in \cite{2015ApJ...805....4Z}. However in this case, we report the LALOs and LATOs in the filament threads exhibiting the failed eruption. Furthermore, we also observe damping in LALOs. The paper is organized as follows: In section 2 we narrate our observation. In section 3 we discuss our data analysis and results. In section 4 we summarize our work and conclusions are drawn.
 

\section{Observation}\label{sec:observation}

\begin{figure}[!ht]
\centering
\includegraphics[scale=0.6,angle=90]{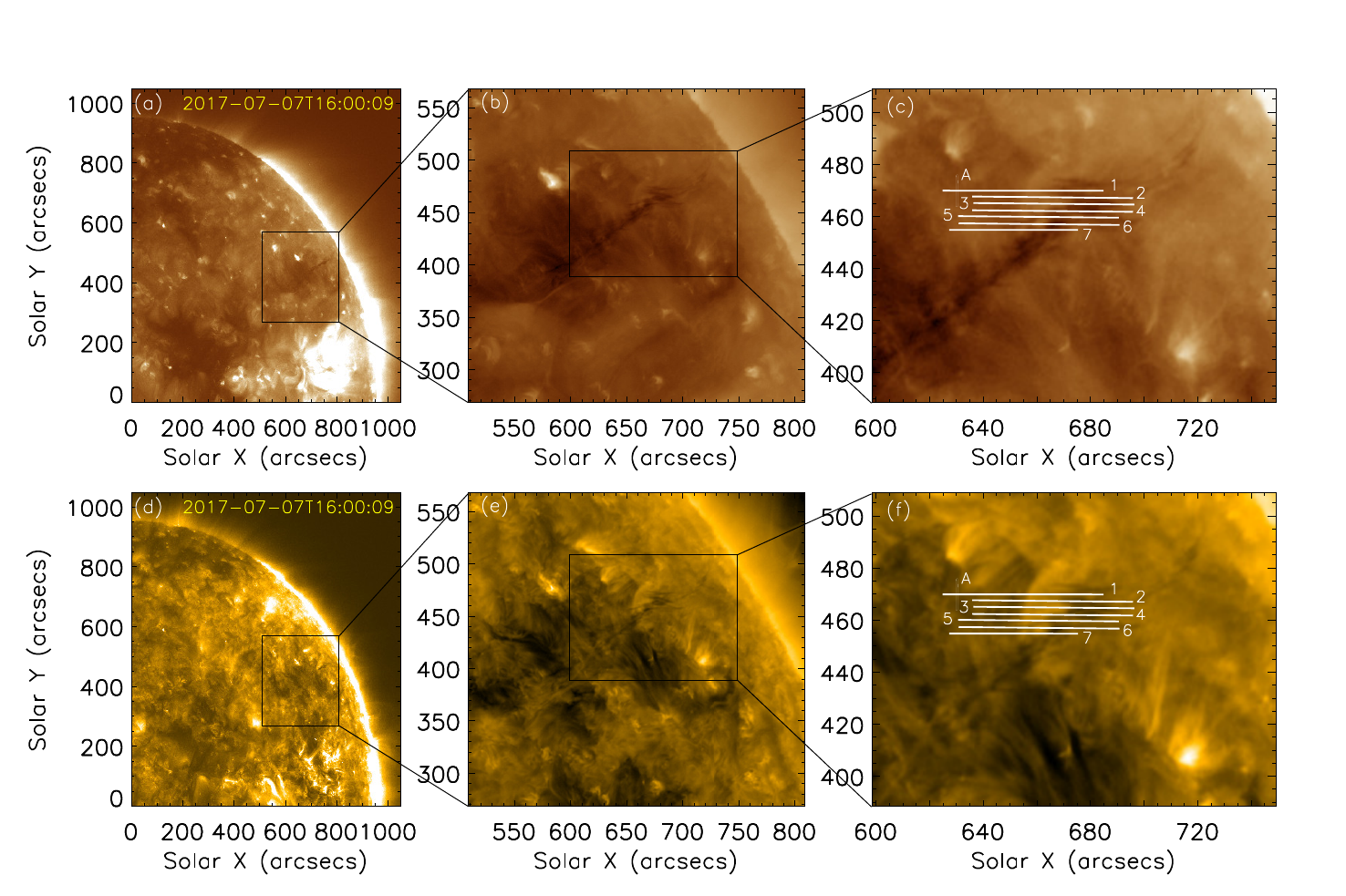}
\caption{The upper row panels and lower row panels shows the filament in {SDO/AIA 193\AA\ and 171\AA\ respectively}. Both panels (a) and (d) show the context of the filament we are considering. Central panels (b) and (e) show a closer view of the studied filament and the region considered in this study (black box). Panels (c) and (f)  depict the filament in AIA 193\AA\ and 171\AA\ in the field of view chosen by the rectangular box in (b) and (e). The artificial slits are overplotted on the filament. The slits are labeled by the corresponding numbers.}
\label{context}
\end{figure}

The event considered in this work occurred in a big filament located in the north-west of the solar disk on 6th July 2017.
In this study we have used the data from the extreme ultraviolet (EUV) passband of Atmospheric Imaging Assembly (AIA) instrument on board Solar Dynamic Observatory \citep[SDO;][]{2012SoPh..275...17L}. AIA provides full disk images of the Sun from seven EUV passbands. In particular we have used the AIA 171~\AA, and AIA 193~\AA \hspace{0.5 mm} passbands. AIA have spatial resolution of 1.2\arcsec \hspace{0.5 mm} with pixel size of 0.6\arcsec and a temporal cadence of 12 seconds.
Figures \ref{context}(a) and \ref{context}(d) shows the north-west quadrant of the solar disk on 7th July 2017 as recorded in AIA 171\AA\ and 193\AA\  respectively. The filament is marked within the black box which is enlarged in Figures \ref{context}(b) and \ref{context}(e).  Figures \ref{context}(c) and \ref{context}(f) shows  a detailed view of the filament. The seven artificial horizontal slits and one artificial vertical slits are overplotted in white in both panels (c) and (f). The horizontal slits are labelled from 1 to 7. The vertical slit is named as slit A. 
We use the same slits to analyze the 171\AA\ and 193\AA\ data.
We check the signatures of oscillations by placing artificial slits co-spatially in AIA 211\AA\ and GONG H$\alpha$ data but found no significant differences. Therefore in the subsequent analysis we focus on 171\AA\ and 193\AA\ passbands of AIA.

\subsection{Filament rise and trigger of the oscillations}
\begin{figure}[!ht]
\centering
\includegraphics[scale=0.6,angle=90]{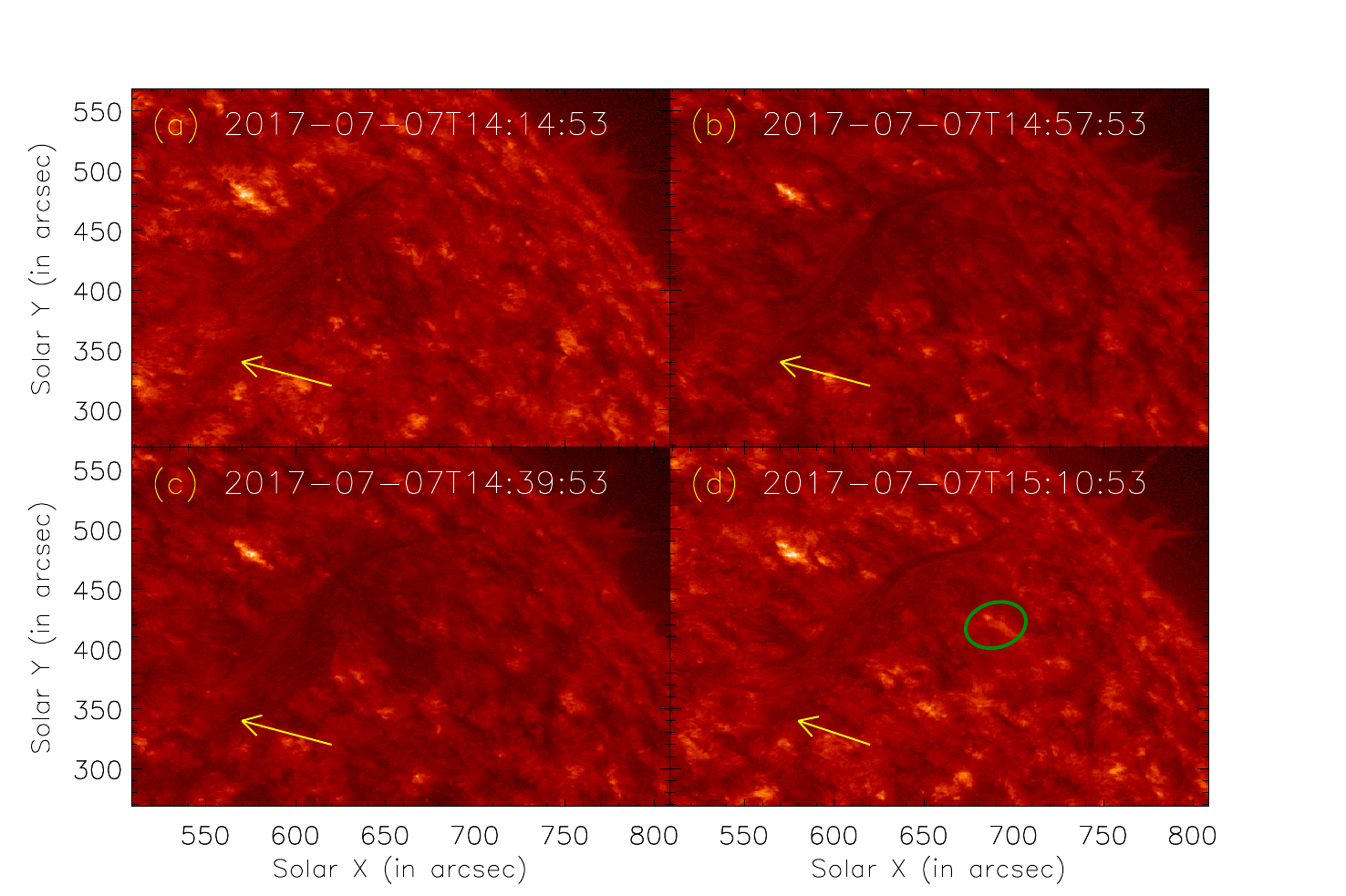}
\caption{Temporal evolution of the rising filament that probably triggered LAOs. ({\it a-c}): Arrows mark the portion of the filament that started rising before LAOs were triggered. {\it d}: Transient brightening follows the lift-off of the filament is outlined by a circle.}
\label{long_time_evolution}
\end{figure}

The event considered in this work consists in a partial and failed eruption of the southern part of the filament and the subsequent LAOs observed in the northern part of the same filament. In this work we focus on the study of the LAOs. It is difficult to study the triggering of the oscillations by the partial eruption because the erupted plasma is very faint and only visible in 304\AA\ channel. We have included an online movie where the faint erupted plasma is seen.
In Figure~\ref{long_time_evolution} and associated animation, it can be clearly seen that the southern part of the filament, marked with the arrows, started to lift-off. The rest of the filament also rise but with a slower rate compared to the southern part. Next, a small portion of the southern part of the filament erupts and transient brightening outlined by a circle in Figure~\ref{long_time_evolution}(d) due to change in magnetic field topology is noted. This brightening is also seen in AIA 171\AA\ and 193\AA\ data (right panels of Figure~\ref{context} at (x,y)=(710,415)). No signatures of coronal mass ejections (CMEs) are seen in LASCO/C2 field-of-view which lead us to conclude that it could be a failed or partial eruption (since only southern portion of the filament exhibits eruption). Furthermore, rest of the filament returned to the equilibrium position after a brief lift-off. This might have triggered LAOs probably due to the large displacement from mean position. It should be noted that this is a mere speculation based on the temporal sequence of the eruption and initiation of the oscillations. In the current study, we focus our analysis on the co-existence of the longitudinal and transverse oscillations and their seismological implications, regardless of how they are triggered.

\subsection{Co-existence of longitudinal and transverse oscillations}
\begin{figure}[ht!]
\centering
\includegraphics[scale=0.6,angle=90]{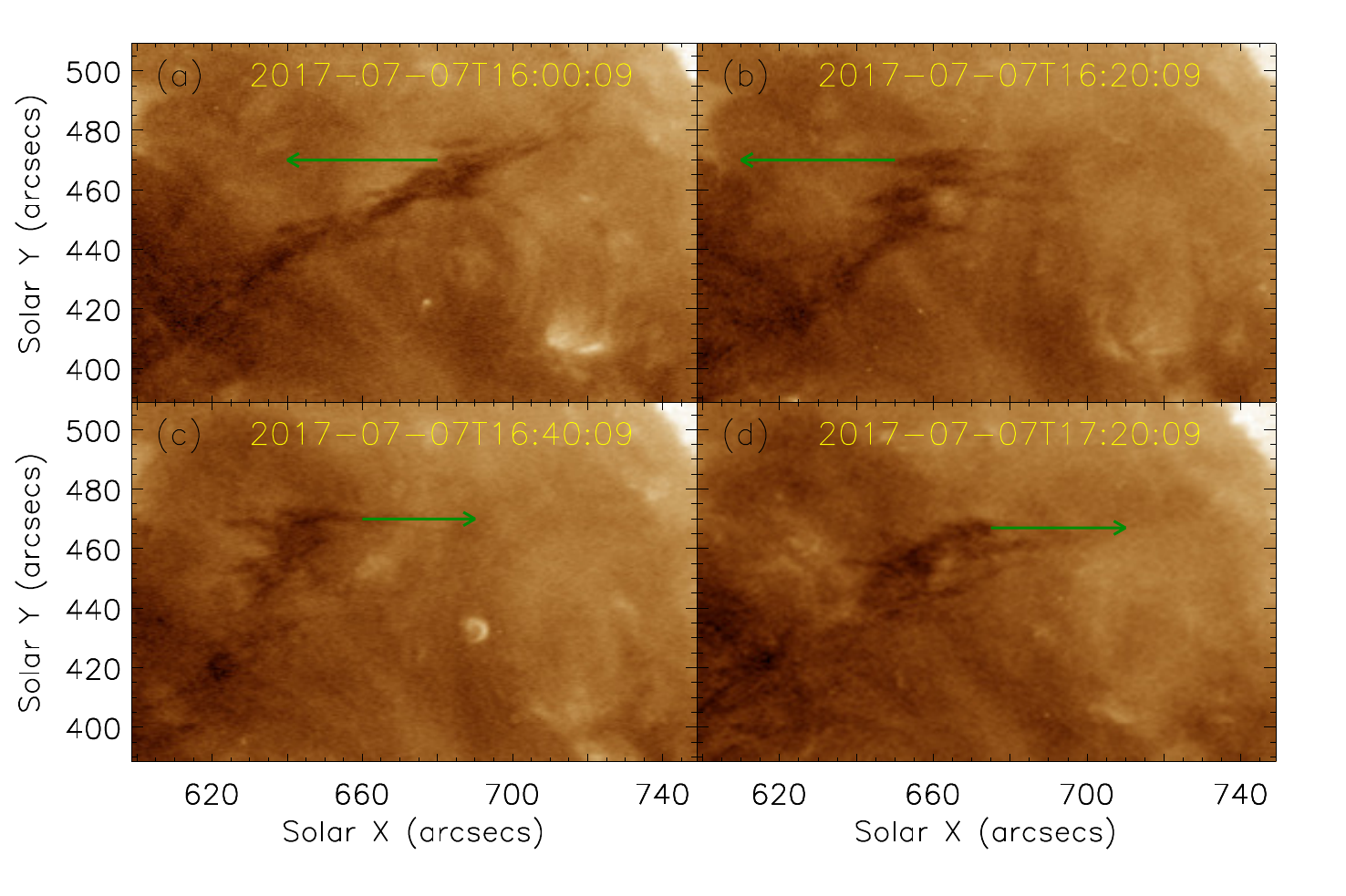}
\caption{Temporal evolution of the filament in AIA 193\AA\ in different phases of the oscillations. The arrows in green represent the direction of the movement of the filament threads at different instances. }
\label{time_evolution}
\end{figure}

Figure \ref{time_evolution}(a) shows the filament at the moment of the triggering of the LAO at 16:00 UT. The green arrow showing the direction of the movement of the filament threads towards the left. In Figure \ref{time_evolution}(b) the movement of filament threads is towards further left direction at 16:20:09 UT. Figure \ref{time_evolution}(c) depicts the rightward movement of the filament threads at 16:40:09 UT. Finally in Figure \ref{time_evolution}(d) the motion is again reversed at 17:20:09 UT.
Furthermore, we also note the signatures of the transverse oscillations in small bunch of filament threads exhibiting longitudinal oscillations. A detailed analysis follows in the coming sections.

\section{Data Analysis and Results}

\subsection{Longitudinal Oscillation in Filament Threads from AIA 171\AA}\label{sec:171longoscil}
\begin{figure}[ht!]
\centering
\includegraphics[scale=0.6,angle=90]{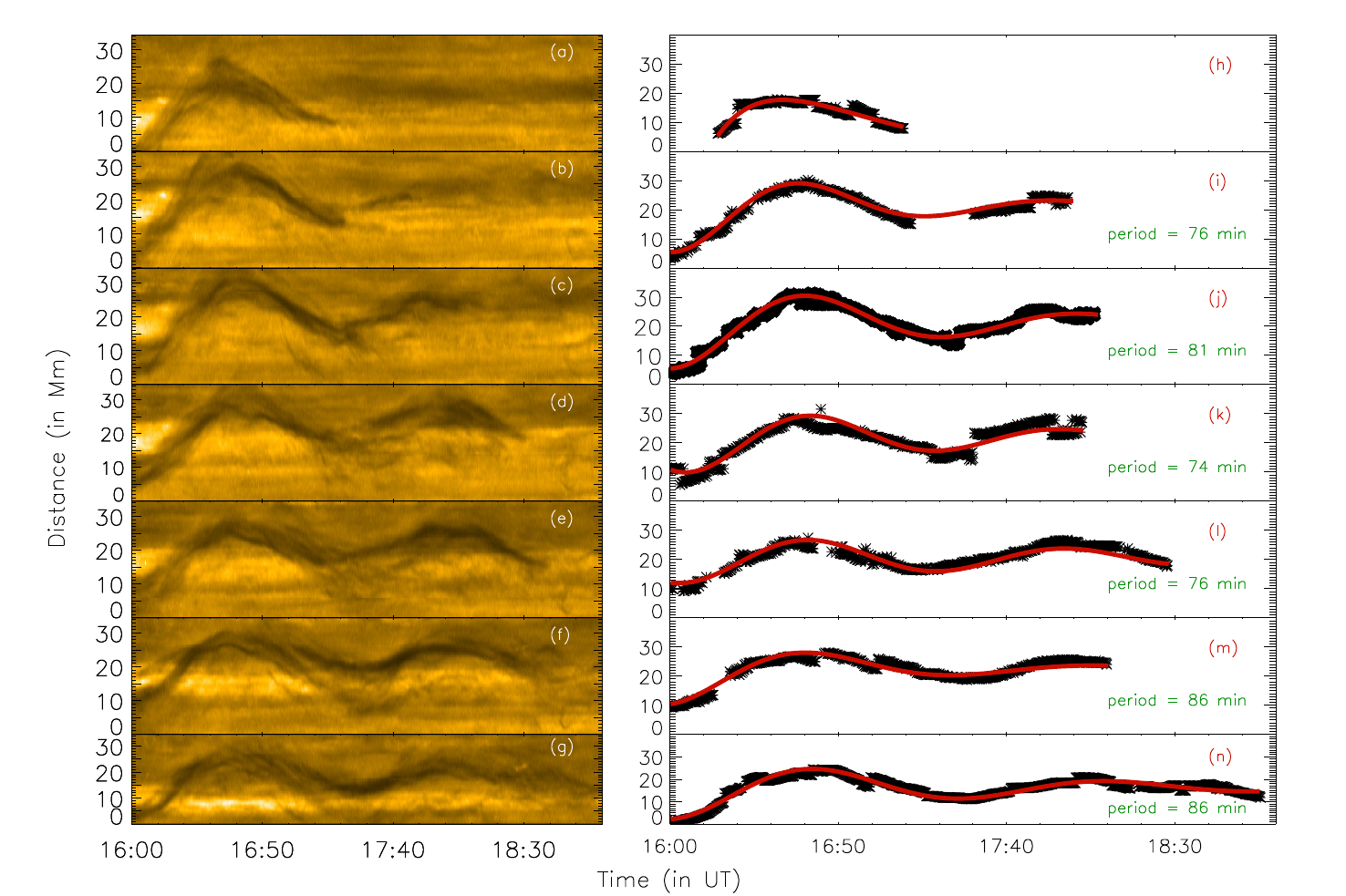}
\caption{Panels (a) to (g) are time-distance diagrams corresponding to the slits 1 to 7 respectively (see Fig. \ref{context}(n). In panels (h) to (n) show the minimum intensity position (black asterisks) of corresponding time distance map of Figures (a)-(h) respectively. The damped exponential sinusoidal fit is depicted by red solid lines. The green dashed lines are drawn to estimate the extent of the damped exponential sinusoidal fit.}
\label{xt_long_all_171}
\end{figure}

As described in Section \ref{sec:observation}, we clearly see the oscillations of filament threads (see available online animation). We identify visually the motion of several threads. We place 7 artificial horizontal slits labeled from 1 to 7 in Figure \ref{context}(c). These slits are placed along the different filament threads to track their motions. These horizontal slits capture the longitudinal oscillations of filament threads. From the artificial slits, we generate time-distance maps, time in the horizontal axis and distance along the slits in the vertical axis ($x-t$ map). Figures \ref{xt_long_all_171}(a)-(g) show the x-t map produced from slits 1-7 respectively. In all the x-t maps we observe longitudinal oscillation of dark filament threads.  In  Figures \ref{xt_long_all_171}(h)-(n), points shown in black are the detected intensity minimum points from corresponding x-t maps of Figures \ref{xt_long_all_171}(a)-(g) respectively. The red curves in the Figures \ref{xt_long_all_171}(h)-(n) are the fitted damped sinusoidal curves to the detected minimum intensity points with best fit parameters.
The damped sinusoidal curve is represented by the following equation \citep[see, e.g.,][]{1999ApJ...520..880A}
\begin{equation}
    y(t)=C+A \sin(\omega t+\phi) \, e^{-t/\tau} \, ,
\label{eq_fitting}
\end{equation}
where C is a constant, $A$ is the amplitude, $\omega$ is the angular frequency, $\tau$ is the damping time and $\phi$ is the initial phase  of the oscillation. We perform the least square fitting in interactive data language (IDL) using MPFIT.pro \citep{2009ASPC..411..251M}. 


The best fit values and the associated errors in the amplitudes, periods ($P={2\pi/\omega}$), damping time and velocity amplitudes of the longitudinal oscillations are quoted in Table \ref{Tab:table_long_171}.

\begin{table}[!ht]
\centering
\caption{Table of parameters for the damped sinusoidal fitting of longitudinal oscillation in filament thread in AIA 171\AA\ }
\small\addtolength{\tabcolsep}{-5pt}
\begin{tabular}{cllll}     
  \hline                   
  Slit No. &\hspace{1mm} A (Mm) & \hspace{1mm} P (min.) & \hspace{1mm} $\tau$ (min.) & \hspace{1mm} $V=\omega \, A (\mathrm{km s^{-1}}$) \\
  \hline
 2 & 16.53 $\pm$ 2.97 & \hspace{1mm} 76 $\pm$ 0.10 & \hspace{1mm} 58 $\pm$ 0.2 & \hspace{1mm} 22.78 $\pm$ 4.12 \\
 3 & 16.09 $\pm$ 2.46 & \hspace{1mm} 81 $\pm$ 0.06 & \hspace{1mm} 71 $\pm$ 0.25 & \hspace{1mm} 20.80 $\pm$ 3.20 \\
 4 & 12.89 $\pm$ 1.79 & \hspace{1mm} 74 $\pm$ 0.06 & \hspace{1mm} 77 $\pm$ 0.28 & \hspace{1mm} 18.24 $\pm$ 2.55 \\
 5 & 8.82 $\pm$ 1.11  & \hspace{1mm} 76 $\pm$ 0.06 & \hspace{1mm} 119 $\pm$ 1 & \hspace{1mm} 12.12 $\pm$ 1.53 \\
 6 & 12.30 $\pm$ 2.16  & \hspace{1mm} 86 $\pm$ 0.14 & \hspace{1mm} 51 $\pm$ 0.24 & \hspace{1mm} 15.04 $\pm$ 2.67  \\
 7 & 14.12 $\pm$ 1.97 & \hspace{1mm} 86 $\pm$ 0.05 & \hspace{1mm} 61 $\pm$ 0.20  & \hspace{1mm} 17.07 $\pm$ 2.39 \\
  \hline
\end{tabular}
\label{Tab:table_long_171}
\end{table}
We discard the data from slit 1 because one sigma errors in time are very large. This large error is because in the slit 1 we only capture one half of a period.

From the table we see that the period ranges from \textit{74} to \textit{86} minutes. This shows a very uniform period values with an average value of \textit{79.83$\pm$0.47} minutes. In contrast, $\tau$ values are more disperse ranging from \textit{51} to \textit{119} minutes. The value of damping time is large in comparison to the value reported in \cite{2009SSRv..149..283T} but comparable to the value reported by \cite{2014ApJ...785...79L}.
The average value of the $\tau$ is estimated to be \textit{72.83$\pm$2.17} minutes. The strength of the damping is measured by the parameter $\tau/P$ that ranges from $\textit{0.59} \pm \textit{0.004} $ to 
\textit{$\textit{1.57} \pm \textit{0.014} $}.
These values indicate a strong damping of the LALOs. The displacement amplitudes, $A$, have also a large range of values from \textit{8} to \textit{16} $\Mm$ and the oscillation velocities, $V=\omega \, A$, range from \textit{12} to \textit{22} $\kms$. The large range of values of $A$ and $V$ indicates that the oscillations in different threads are excited with different energies during the triggering process. 

\subsection{Longitudinal Oscillation in Filament Threads from AIA 193~\AA}
\begin{figure}[!ht]
\centering
\includegraphics[scale=0.6,angle=90]{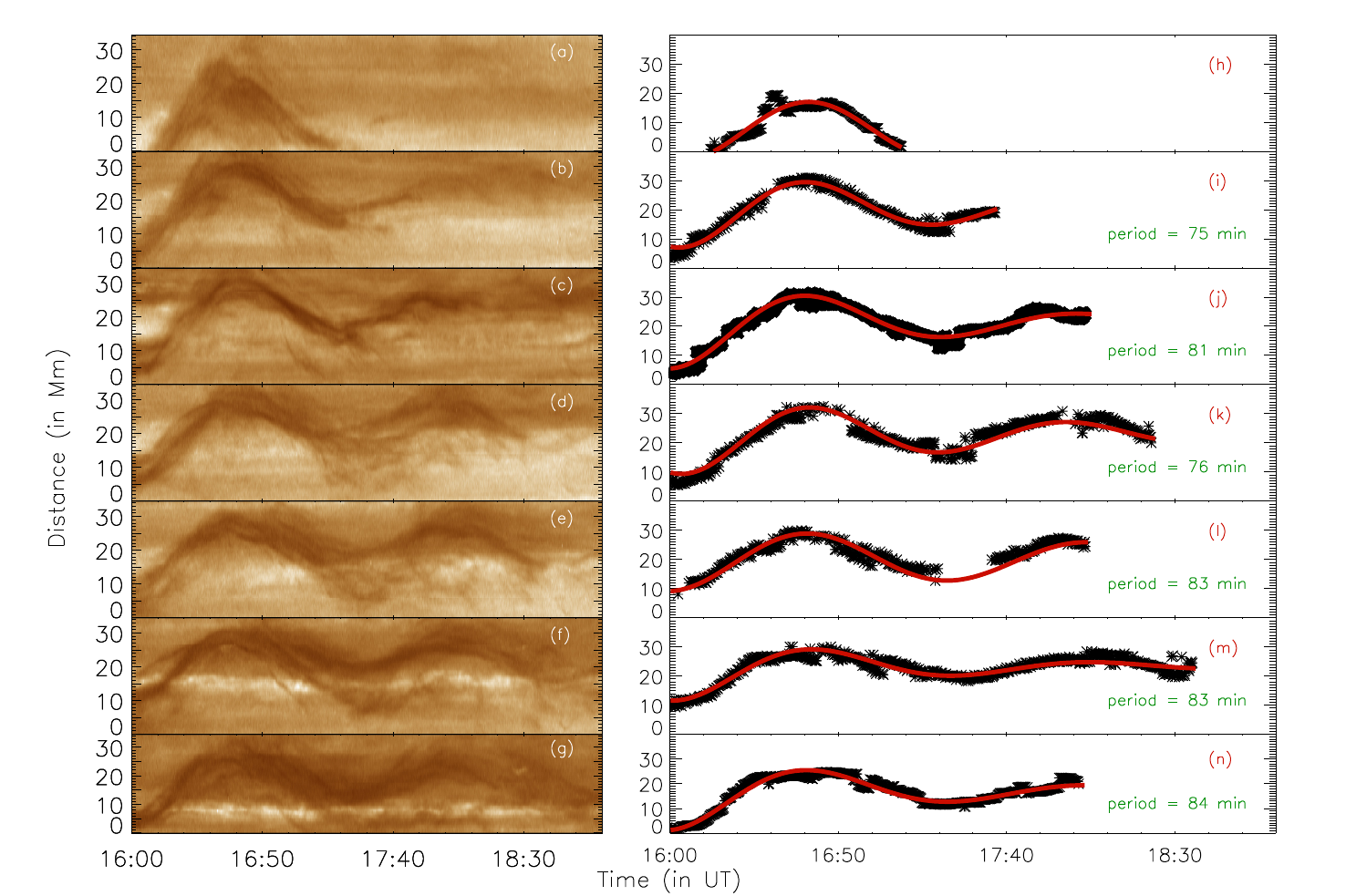}
\caption{The subfigure (a)-(g) are time distance map of slits 1-7 (see upper right panel of Figure \ref{context}) respectively in AIA 193\AA. In Figures (h)-(n) we the black asterisks represent the minimum intensity points of corresponding time distance map of Figures (a)-(h) respectively. The damped exponential sinusoidal fit is depicted by red solid lines. The green dashed lines are drawn to estimate the extent of the damped exponential sinusoidal fit.}
\label{xt_long_all_193}
\end{figure}

We additionally analyze the AIA~193~\AA\ channel using the same 7 horizontal slits shown in Figure \ref{context}(c). Figure \ref{xt_long_all_193} shows the seven time-distance diagrams from the AIA 193~\AA\ data. As in previous case black asterisks show the position of the minimum intensity associated to the central location of the threads where the absorption is maximum. We fit the oscillation with Equation (\ref{eq_fitting}) and the resulting parameters are shown in the Table \ref{Tab:table_long_193}. The time period varies from \textit{75} minutes to \textit{84} minutes with an average value of \textit{80.3$\pm$0.4} minutes. The damping time varies in a wide range from \textit{62} minutes to \textit{203} minutes and the average value of the damping time is estimated to be \textit{97.67$\pm$2.74} minutes. The displacement amplitudes, $A$, have also a large range of values from \textit{10} to \textit{16} $\Mm$. As in previous case we discard the fitting parameters from slit 1. The results are similar to the findings from the AIA 171 wavelength. The similarity of results are expected as the geometry remains same for two different wavelengths. 

\begin{table}[!ht]
\centering
  \caption{Table of parameters for the damped sinusoidal fitting for longitudinal oscillation of filament thread in AIA 193\AA\ }
\small\addtolength{\tabcolsep}{-5pt}
\begin{tabular}{ccccc}     
  \hline                   
 Slit No. & A (Mm) & \hspace{1mm} P (min.) & $\tau$ (min.) & \hspace{2mm} $V=\omega \, A (\mathrm{km s^{-1}}$) \\
\hline
slit 2 &  14.13 $\pm$ 2.02 & \hspace{1mm} 75 $\pm$ 0.06 & \hspace{1mm} 88 $\pm$ 0.43 &  19.72 $\pm$ 2.83 \\
slit 3 &  16.09 $\pm$ 2.46 & \hspace{1mm} 81 $\pm$ 0.07 & \hspace{1mm} 71$\pm$ 0.25 &  20.91 $\pm$  3.22 \\
slit 4 &  14.21 $\pm$ 1.89 & \hspace{1mm} 76 $\pm$ 0.05 & \hspace{1mm} 97 $\pm$ 0.04 &  19.55 $\pm$  2.61 \\
slit 5 &  10.89 $\pm$ 1.33 & \hspace{1mm} 83 $\pm$ 0.04 & \hspace{1mm} 203 $\pm$ 1.50 &  13.73 $\pm$ 1.68 \\
slit 6 &  12.00 $\pm$ 1.82   & \hspace{1mm} 83 $\pm$ 0.10 & \hspace{1mm} 62 $\pm$ 0.29 &  15.20 $\pm$ 2.32\\
slit 7 & 16.01 $\pm$ 2.56  & \hspace{1mm} 84 $\pm$ 0.08 & \hspace{1mm} 65 $\pm$ 0.23  & 20.04 $\pm$ 3.22 \\
  \hline
\end{tabular}
\label{Tab:table_long_193}
\end{table}

The parameters of the longitudinal oscillations are in agreement with the previous findings \citep{2014ApJ...785...79L,2017ApJ...842...27Z} with similar range of periods and damping times. We see little differences in the fitted parameters in AIA 171\AA\ and 193\AA\ passbands. Since we are following the cool plasma, the absorption could be slightly different in different channels of AIA which influences the estimation of the best fit parameters.

\subsection{Transverse Oscillation in Filament Threads}\label{sec:transverseoscil}
\begin{figure}[!ht]
\centering
\includegraphics[scale=0.6,angle=90]{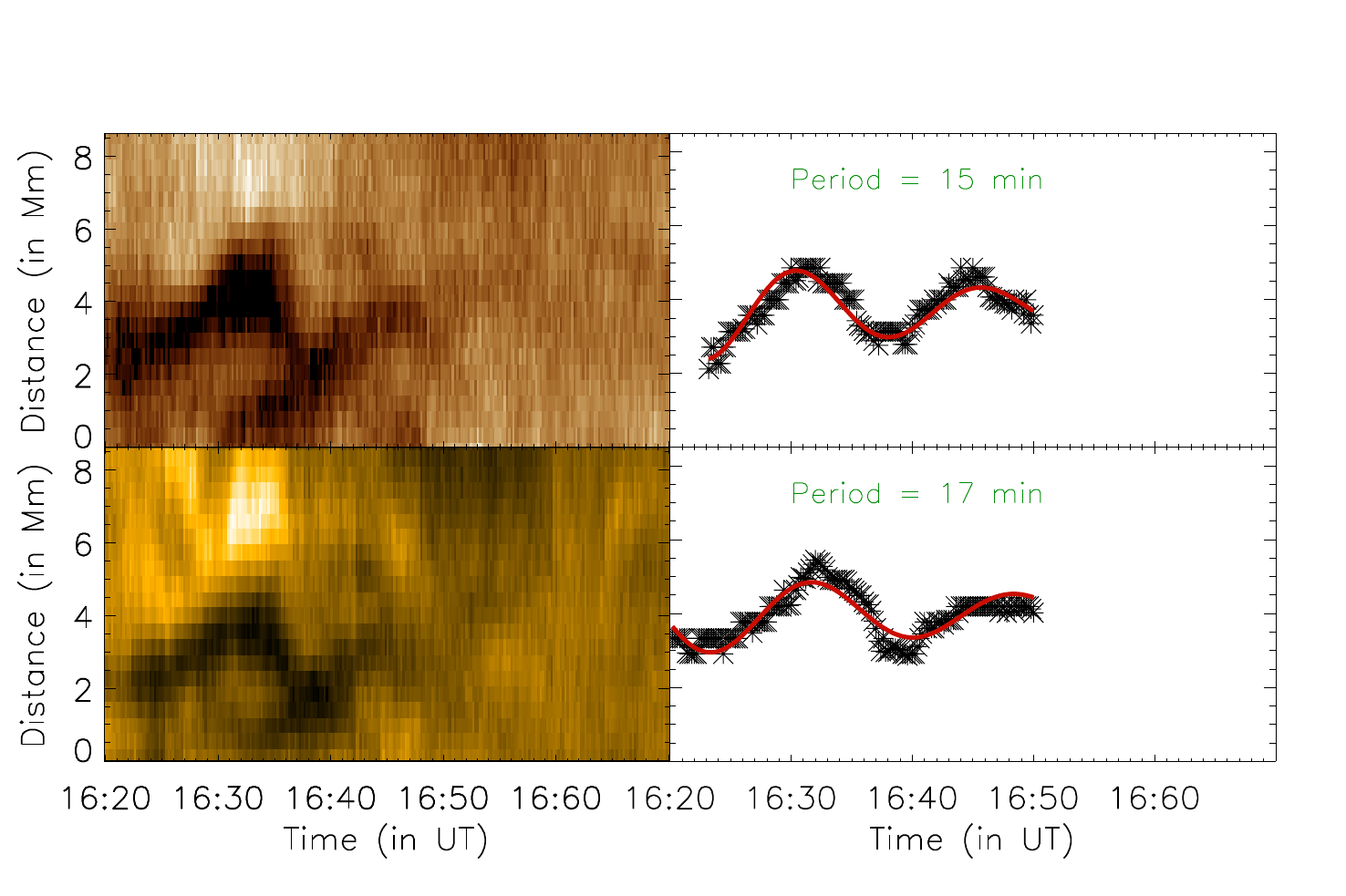}
\caption{Upper left panel shows the time-distance map obtained using artificial slit A (see lower right panel of Figure \ref{context}) in AIA 193\AA\ data. Lower left panel is similar to the upper left but for AIA 171\AA\. Black asterisks in upper (and lower) right panel represent the minimum intensity corresponding to time-distance map shown in the left panels. Overplotted in red color is the best fit exponentially damped sinusoidal curve to the minimum intensity points.}
\label{xt_trans1}
\end{figure}

The uppermost thread of the filament in slit 7 also oscillates in the transverse direction for a shorter time span in comparison to longitudinal oscillation. Tracking the transverse motion is difficult due the simultaneous longitudinal displacement of the filament thread. We place a new slit labeled as A perpendicular to slit 7 (see Figs. \ref{context}(c) and (f)). We move the slit A along the slit 1 following the longitudinal movement. In this way we can track the motion in the transverse direction producing the time-distance diagram in the slit A using the AIA 171\AA\ and 193~\AA. It should be noted that the slit A remains perpendicular to the slit 1 at all instances so that spurious oscillations are not detected. From the diagrams in both channels we identify a signature of the oscillation (see Figures \ref{xt_trans1}). As in the case of the longitudinal oscillations we fit the decaying sinusoidal function (Eq. (\ref{eq_fitting})) to these intensity minimum points.

The parameters of oscillation are shown in Table \ref{table_trans}.
\begin{table}[!ht]
\centering
\caption{Table of parameters of sinusoidal fitting for transverse oscillation in the filament thread.}
\label{table_trans}
\small\addtolength{\tabcolsep}{-5pt}
\begin{tabular}{ccccc}     
  \hline                   
Wavelength Channel & A (Mm) & P (min.) & $\tau$ (min.) & $V=\omega \, A (\mathrm{km s^{-1}}$) \\
  \hline
AIA 171 & 1.17 $\pm$ 0.34 &  \hspace{1mm}17 $\pm$ 0.04 & 21 $\pm$ 9 & 7.21 $\pm$ 2.11 \\
AIA 193 & 1.54 $\pm$ 0.49 & \hspace{1mm} 15 $\pm$ 0.04 & 27 $\pm$ 6  &  10.75 $\pm$ 3.45 \\
  \hline
\end{tabular}
\end{table}
The transverse oscillations have much smaller amplitudes with displacements of \textit{1.2} and \textit{1.6} Mm in 171~\AA\ and 193~\AA\ respectively. In contrast, the velocities are comparable to the longitudinal case. The period of oscillations are also considerably smaller than the longitudinal motions. As we discuss in the following section the origin of the transverse oscillation is magnetoacoustic and the restoring force is associated with the magnetic field. The damping is comparable to the period indicating strong damping of the transverse oscillations. However, the measurement of the damping time is not reliable because the filament threads disappear while exhibiting transverse motions as seen in figure~\ref{xt_trans1}. Transverse oscillations with periods between \textit{3} to \textit{5} minutes in the filament threads are reported in previous studies \citep{2009ApJ...704..870L,2015ApJ...809...71O}. In this work, we found long periods of about \textit{15} minutes. Since period of transverse oscillations depends on the length of the fluxtube, long periods in our study indicate that the filament threads might be longer. We discuss this in detail in the next section.

\section{Seismology}\label{sec:seismology}
\begin{figure}[!ht]
\centering\includegraphics[scale=0.4,angle=0]{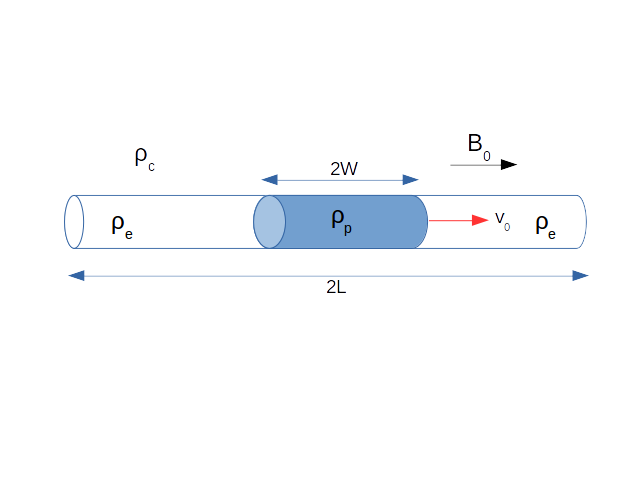}
\vspace{-3cm}
\caption{Schematic diagram of the magnetic field and plasma configuration. The blue color smaller cylinder represent the flowing thread inside the  magnetic tube which is represented by the bigger white cylinder. The length of the magnetic tube is 2L and the length of the filament thread is 2W. The density of the thread is $\rho_{p}$, the density of the magnetic tube is $\rho_{e}$ and the density of the outer corona is $\rho_{c}$. The thread is moving along  axis of the tube with velocity $v_{0}$. }
\label{image_cylinder}
\end{figure}
In this section, we combine the results from our observations with the theoretical models to infer parameters of the prominence. For the longitudinal oscillations, we use the pendulum model by \citet{2012ApJ...750L...1L} to estimate the radius of curvature of the filament. In the pendulum model, it is assumed that the motion of the plasma is mainly along the dipped field lines that support the prominence against gravity. In these conditions, the gravity is the dominant restoring force and the period depends exclusively on the radius of curvature of the dips supporting the cool prominence plasma, $R$ as
\begin{equation}
\omega = \frac{2\pi}{P} = \sqrt{ \frac{g_{0}}{R} } \, ,
\label{eq_rad}
\end{equation} 
where $\omega$ is the angular frequency, $g_{0}$ is the surface gravity of the Sun, P is the period of the longitudinal oscillation. Using the average value of the period for 171~\AA\ data, $P=\textit{79.83}\pm\textit{0.47}$ minutes we can estimate, $R=160 \pm 1 \Mm$.
Similarly, using the average periods from the 193~\AA\ data, $P=\textit{80.3}\pm \textit{0.4}$ we obtain an identical result, $R=161\pm1$. Both results are similar because the geometry of the filament is same and independent of the AIA wavelength.


Following \citet{2012ApJ...750L...1L} and \citet{2014ApJ...785...79L} we can estimate the minimum magnetic field by assuming that the magnetic tension in the dipped part of the tubes must be larger than the weight of the threads. With this assumption it is possible to find the relation,
\begin{equation}
B(G) \geq 26 \left( \frac{n_{e}}{10^{11}}\right) ^\frac{1}{2}  P(hr) \, ,
\label{eq_mag_long}
\end{equation}
where B is the magnetic field, $n_{e}$ is the electron number density in cm$^{-3}$ and $P$ is the period of the oscillation in hours. In absence of a direct measurement, here we use the typical value of electron number density in the range of $10^{10}$ - $10^{11}$ cm$^{-3}$  \citep{2010SSRv..151..243L}. In the estimation of magnetic field the density is the most important source of the uncertainty as the error in estimation of period is much less than the range of density given. Under these conditions the equation \ref{eq_mag_long} becomes
\begin{equation}
    B(G) > (17 \pm 9) P(hr) \, .
\end{equation}
Using this average value of the longitudinal oscillations we obtain the lower limit of the magnetic field of 22.6$\pm$ 11.9 G  and 22.8$\pm$12.0 G for 171~\AA\ and 193~\AA\ respectively which are comparable with the typical values of reported magnetic field \citep{2010SSRv..151..333M,2010SSRv..151..243L} by direct means. 

In contrast to the LALOs, the transverse oscillations of Section \ref{sec:transverseoscil} have a MHD nature. So we can study the transverse oscillation applying MHD models. We apply the linear model by \citet{2002ApJ...580..550D,2005SoPh..229...79D}. Figure \ref{image_cylinder} shows the equilibrium configuration we consider. The prominence thread has a length $2W$ and density $\rho_{p}$. The thread is embedded in a flux tube of length $2L$ and density $\rho_{e}$. Both sides of the flux tubes are called evacuated regions and they have probably a density larger than the ambient corona \citep[see models by][]{2005SoPh..229...79D,2002ApJ...580..550D}. The piecewise density along the tube is given by
\begin{equation}\label{eq:piecewisedens}
    \rho_0 (z,t) = 
    \begin{dcases}
       \; \rho_{p} & |z| \leqslant W \, ,\\
       \; \rho_{e} & |z| > W \, ,
    \end{dcases}
\end{equation}
where $z=0$ is the middle point of the tube and $z = \pm L$ are the end points of the cylinder. The magnetic field $\vec{B}$ is along the axis of the cylinder and it is same inside and outside of the cylinder. To study the dynamics of the thread we can assume the ratio of the width of the thread to length of it is very small which is called thin tube approximation. \cite{2008ApJ...678L.153T} performed a similar seismological analysis. They found that the flow has a little influence on the oscillation periods with corrections below the error bar of their observation.
For this reason we have only studied the special case of negligible flow of the thread. The normal modes of steady non flowing filament threads were studied by \citet{2002ApJ...580..550D} and \citet{2005SoPh..229...79D}. The system supports fast and Alfv\'en waves. The Alfv\'en waves results azimuthal (i.e. torsional) motions so they do not cause displacement of the cylinder axis. In contrast, the kink fast mode produces displacement of cylinder axis resulting the transverse oscillation of filament thread as we observe. \cite{2005SoPh..229...79D} gives a simple dispersion relation since it assume the thin tube approximation. The following formula (Eq. (27) in their study) gives the dispersion relation
\begin{equation}\label{eq_disprsn}
    \tan \left( \Omega (1-l) \sqrt{\frac{1+e}{2}} \right) - \sqrt{\frac{1+e}{1+c}} \cot \left(\Omega l \frac{1+c}{2}\right)=0 \, ,
\end{equation}
where $e=\rho_{e}/\rho_{c}$, $c=\rho_{p}/\rho_{c}$, $l=W/L$, and $\Omega=\omega L /v_{Ac}$, with $\omega $ the frequency and ${v_{Ac}}$ the coronal Alfv\'en speed. The Alfv\'en speed in the thread plasma and in the evacuated parts of the cylinder are respectively $v_{Ap}=c^{-1/2} v_{Ac}$ and $v_{Ae}=e^{-1/2} v_{Ac}$ expressed in terms of the density ratios $c$ and $e$ and $v_\mathrm{A c}^{2} =B^{2}/(\mu \rho_{c}) $. The frequency of kink mode is given by smallest positive root of Equation (\ref{eq_disprsn}).  This equation establish a relation between the field strength $B$, the period $P$ of the transverse oscillation, density contrasts $c$ and $e$, thread half-length $W$ and half-length of the field line $L$. It is usual to compute the magnetic field measuring some parameters or assuming their typical values. However, we use the values of $B$ obtained from the seismology of the longitudinal oscillations and then we can infer the value of the length of the field line $2L$. This is a completely new approach combining both polarizations. From Equation (\ref{eq_disprsn}) we obtain the half-length of the flux tube in terms of the rest of parameters,
\begin{equation}\label{eq_length}
    L= W + \frac{v_{Ac}}{\omega} \sqrt{\frac{2}{1+e}} \tan^{-1} \left[ \sqrt{\frac{1+e}{1+c}} \cot \left(\frac{\omega W}{v_{Ac}} \sqrt{\frac{1+c}{2}}\right) \right] \, .
\end{equation}
We use the average period of the transverse oscillations, $P=\textit{16}$ minutes (see Table \ref{table_trans}). From Figure \ref{xt_trans1} we estimate the thread length to $2 W \sim 1.5 \Mm$. From the longitudinal oscillations seismology we have estimated $B \sim 11 - 35$ G using the pendulum model. Assuming typical coronal electron densities of $10^{8}- 10^{9}  \mathrm{cm}^{-3}$ we obtain that $v_{Ac} \geq 759 \kms$ up to several thousands of $\kms$. However, we assume a maximum value of $1500\kms$ in agreement with the observations. We assume typical values for $c= 100 - 200$ \citep[see][]{2010SSRv..151..333M,2010SSRv..151..243L} and $e$ identical to the range of values considered by \cite{2005SoPh..229...79D}, i.e., $e = 0.6 - 2$. With these ranges of parameters we obtain a range of lengths of the field lines as $2L= 206 - 713 \Mm$. The range of lengths comprises long to very long field lines.  There is no direct measurement of the length of the field lines. However, theoretical models assume that the field lines are long with few hundreds of Mm \citep[see][]{2010SSRv..151..333M,luna2012,terradas2008} and our range of values overlaps the theoretical estimations. Probably, the largest values of the range obtained here by seismology are overestimated due to the uncertainties of the method.
This result agrees with the idea that field lines supporting the prominence are very large. Many models of prominence field structure assumes a flux rope with many turns. We can assume that the radius of the rope is $R$. Then the ratio $L/\pi R$ gives an idea of the number of turns of the flux rope. From the seismology we obtain a number smaller than one suggesting that the field structure consists in a rope with less than one turn.

The damping of the oscillation offers additional information about the prominence and the processes involved on it. Similarly to the oscillation period, assuming a physical damping mechanism we can infer some parameters. In \citet{luna2012} and later in \citet{2016A&A...591A.131R} we modelled the damping as a result of the mass accretion associated to the evaporation-condensation process \citep{2010SSRv..151..333M}. From the approximate Equation (46) of \citet{2016A&A...591A.131R} we can estimate the mass accretion rate as
\begin{equation}\label{eq:rudermanlunadamping}
    \frac{\Dot{M}}{M} = \frac{2.12}{\tau} \, ,
\end{equation}
being $M$ the mass of the thread and $\Dot{M}=dM/dt$. The average longitudinal damping time is $\tau=85 \min$. Using this number we can estimate that $\Dot{M}/M=0.025 \, \mathrm{minutes}^{-1}$. However, this number should be considered as a crude estimation because Equation (46) from \citet{2016A&A...591A.131R} is an approximation.
\begin{figure}[!ht]
\centering
\includegraphics[scale=0.6,angle=90]{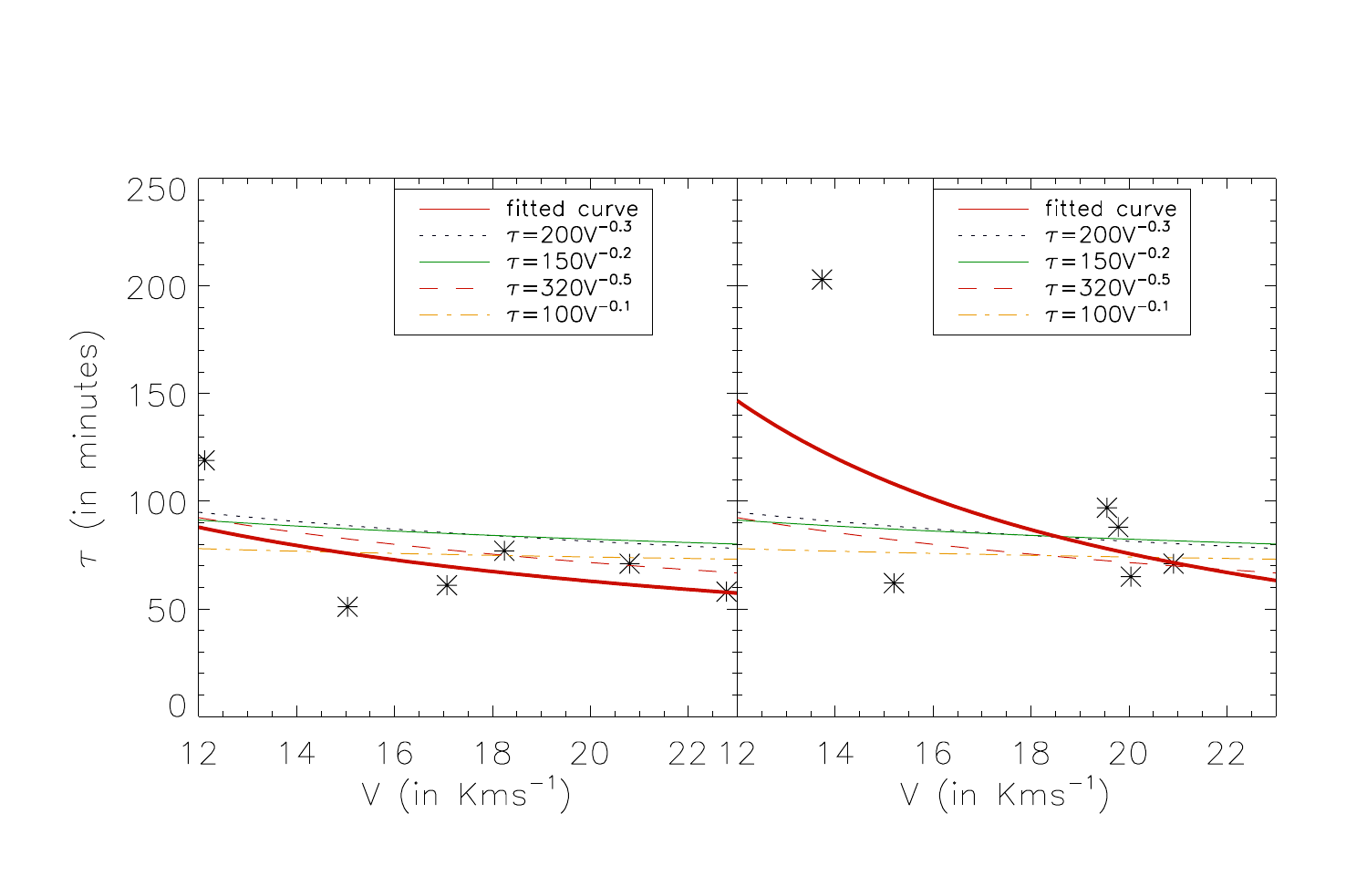}
\caption{ Left panel shows the variation of damping time ($\tau$) with velocity (v) of the oscillation in AIA 171\AA. Right panel is similar to the left but for AIA 193\AA. The black asterisks represents the observed data. A non-linear curve given by the relation $\tau \sim a~V^b$ is fitted to the data. Curves with different exponents and constants are overplotted in different colors and linestyles. Curve in solid red in left (right) panel represents the best fit curve with $a$ and $b$ to be 450 (3657) and -0.7(-1.3) respectively.}
\label{tau_vs_v}
\end{figure}

\citet{luna2018} found observationally a non-linear behaviour of LAOs. The authors found that the damping time $\tau$ depends on the oscillation velocity, $V$. In Figure \ref{tau_vs_v} we have plotted $\tau$ as function of the velocity amplitude on each slit. In both panels it seems that there is a trend of reducing $\tau$ for increasing $V$ in agreement with previous findings. \citet{2013A&A...554A.124Z} found that the radiative cooling could explain this trend and consequently the damping. The authors found a nonlinear relationship between $\tau$ and $V$ as $\tau \sim V^b$ with $b=-0.3$. In Figure \ref{tau_vs_v} we have over plotted this scaling law. The Zhang's scaling law seems to follow the trend of the data. However, different exponents of the scaling law, $b$ produces similar results. In addition, we have fitted the data with a general form $\tau= a \, V^b$ and we have obtained that $b \sim -1.3$ that is very different of the Zhang's scaling law. Moreover, Zhang et al. found a complex dependence of $\tau$ of the different parameters of the thread and the field line as the thread length, field line length, the geometry of the dip and the velocity amplitude. The exponents of the parameters are quite large expect for $V$ indicating that $\tau$ is more sensitive to the parameters of the thread and the field lines than for $V$. In the model by \citet{2016A&A...591A.131R} the damping is not related directly to the oscillation velocity. However, events with larger $V$ are associated with violent events, which could produce increased evaporation and consequently stronger damping. More recently, \citet{Zhang:2019ug} have found by numerical simulations that wave leakage is an important ingredient of the LAOs damping but in weak field prominences. It is necessary additional observational and theoretical efforts to determine the combination of mechanisms responsible of the strong LAOs damping and their applicability to seismology.

\section{Summary and Conclusions}
In this work, we report simultaneous longitudinal and transverse oscillation in the same filament threads for the first time. The event occurred in a large filament located in the North-West quadrant of the solar disk on 7th July 2017. The triggering of the LAOs is associated to the failed eruption of the filament. The filament starts to rise around 13 UT. 
The eruption fails and the filament falls to the equilibrium position. This produces the oscillations of part of the cool plasma of the filament.
The oscillations are mainly longitudinal but one thread also oscillates in the transverse direction to the local field. We use the time-distance technique to analyze the oscillations and we obtain the oscillation parameters in both AIA 171\AA\ and 193\AA\ channels. Both channels show identical oscillations. The periods obtained are in the range of \textit{75} to \textit{86} minutes with an average value of \textit{80} minutes. The velocities of the oscillations are in the range of \textit{12} to \textit{23} $\kms$ being LALOs. The oscillations are strongly damped with damping times comparable to the oscillation period. The transverse oscillations in one of the threads have a considerably smaller period of \textit{16} minutes, an average velocity amplitude of \textit{9} $\kms$in both channels, and an average damping time of \textit{24} minutes.

We have applied seismological technique to both oscillation polarizations. For the longitudinally oscillating case, we have used the pendulum model and we determined the radius of curvature of the magnetic dips hosting the prominence, $R$, and the minimum field strength, $B$, required to support the mass against gravity. The resulting average values in the 7 slits are $R=160\Mm$ and $B=23 \pm 12 $ G. For the transverse oscillations the restoring force is associated to the magnetic Lorentz force. In a novel way we combine the magnetic field obtained with the longitudinal oscillations with the dispersion relation of the transverse modes to obtain the length of the field lines hosting the prominence. The length of these field lines are in the range of 206 to 713 Mm. 

We have applied the seismological techniques in a novel way that has allowed to measure the strength, the curvature of the dips and the length of the magnetic field lines supporting the prominence. Our findings suggest that the prominence structure consists in a flux rope with few turns, probably less than one.\\\\

{\bf Acknowledgements}\\
We thank referee for valuable comments. VP was supported by the GOA-2015-014 (KU~Leuven) and the European Research Council (ERC) under the European Union's Horizon 2020 research and innovation programme (grant agreement No 724326). M. L. acknowledges the support by the Spanish Ministry of Economy and Competitiveness (MINECO) through projects AYA2014-55078-P and under the 2015 Severo Ochoa Program MINECO SEV-2015-0548.  M.L. \& V.P. acknowledge support from the International Space Science Institute (ISSI), Bern, Switzerland to the International Team 413 ``Large-Amplitude Oscillations as a Probe of Quiescent and Erupting Solar Prominences'' (P.I. M. Luna). DB and VP acknowledge the support from the ISSI-Beijing for the formation of the international team  titled "The eruption of solar filaments and the associated mass and energy transport” and its activities. R.M. acknowledges the support of United Grants Commission (UGC) and Center of Excellence in Space Sciences India (CESSI), IISER KOLKATA.  \\\\


\end{document}